\documentclass[letterpaper, 10 pt, conference]{ieeeconf}  

\IEEEoverridecommandlockouts                              

\usepackage{graphics} 
\usepackage{epsfig} 
\usepackage{amsmath} 
\usepackage{amssymb}  
\let\labelindent\relax
\usepackage{enumitem}
\usepackage{multirow}
\usepackage{balance}
\usepackage{array}
\usepackage[ruled,vlined]{algorithm2e}
\usepackage{bm}

\newcommand{\mbf}[1]{\mathbf{#1}}

\newcommand{\mc}[1]{\mathcal{#1}}

\usepackage[deletedmarkup=sout,authormarkup=superscript]{changes}
\usepackage{xcolor}
\definechangesauthor[name={Sophie}, color=blue]{SH}
\definechangesauthor[name={Enno}, color=green]{EB}
\definechangesauthor[name={Bepe}, color=blue]{GB}

\makeatletter
\let\NAT@parse\undefined
\makeatother
\usepackage[hidelinks]{hyperref}  

\title{\LARGE \bf
	Carbon-Aware Computing in a Network of Data Centers:\\
	A Hierarchical Game-Theoretic Approach
}

\author{Enno Breukelman$^{1}$, Sophie Hall$^{2}$, Giuseppe Belgioioso$^{2}$, and Florian Dörfler$^{2}$
	\thanks{$^{1}$Enno Breukelman is with the KTH Royal Institute of Technology, School of Electrical Engineering and Computer Science, Division of Decision and Control Systems, Malvinas väg 10, SE-100 44 Stockholm, Sweden. {\tt\small cebre@kth.se}}%
	\thanks{$^{2}$Sophie Hall, Giuseppe Belgioioso, and Florian Dörfler are with the Automatic Control Laboratory, Department of Electrical Engineering and Information Technology, ETH Zurich, Physikstrasse 3, 8092 Zürich, Switzerland. {\tt\small \{gbelgioioso, shall, dorfler\}@ethz.ch}}
	\thanks{This work is supported by the SNSF via NCCR Automation (Grant Number 180545).}
	}%

\begin{document}

\maketitle
\thispagestyle{empty}
\pagestyle{empty}

\begin{abstract}
Over the past decade, the continuous surge in cloud computing demand has intensified data center workloads, leading to significant carbon emissions and driving the need for improving their efficiency and sustainability.
This paper focuses on the optimal allocation problem of batch compute loads with temporal and spatial flexibility across a global network of data centers.
We propose a bilevel game-theoretic solution approach that captures the inherent hierarchical relationship between supervisory control objectives, such as carbon reduction and peak shaving, and operational objectives, such as priority-aware scheduling. 
Numerical simulations with real carbon intensity data demonstrate that the proposed approach successfully reduces carbon emissions while simultaneously ensuring operational reliability and priority-aware scheduling.
	
\end{abstract}


\section{INTRODUCTION}

Between 2010 and 2020, the global compute load on data centers (DCs) has increased more than 9 times \cite{InternetTraffic2020}. As a consequence, the number of hyper-scale DCs has doubled between 2015 and 2021, reaching 700 installed facilities worldwide \cite{Synergy2021}.
For their operation, DCs require a significant supply of electricity from the grid, about 1-1.5\% of the global electricity demand in 2022, corresponding to 230-340 TWh \cite{IEA2023}. 
Unfortunately, the electricity they run on is still coming from predominantly carbon-intensive sources. Incorporating more renewable energy into the electricity mix is challenging due to supply variability, which depends on both time of day and geographical location.
Remarkably, a large share of the global compute load is not time-sensitive, allowing a delayed execution, nor bound to a specific DC, allowing it to be executed at a different DC.
Therefore, various compute-demand management and real-time routing mechanisms exploit this flexibility to mitigate the carbon impact and provide ancillary services to the power grid.

Most of the existing work focuses on real-time routing of temporally inflexible compute jobs \cite{Khosravi2017}, such as on-demand services and search engine requests. Furthermore, loads with either spatial or temporal flexibility \cite{Zhenhua2013,Rahmani2022} are considered, but rarely the combination of both. 
Compute jobs are also mostly considered part of an aggregate compute load, which does not differentiate between individual jobs \cite{Minxian2020}.
Typical objectives considered by these allocation mechanisms include reducing monetary expenses, carbon taxes \cite{BuchbinderYear,Khosravi2017}, carbon emissions \cite{Shuja2016}, and inducing peak shaving \cite{Zhenhua2013,Dabbagh2019}.
Additional modeling features investigated in existing research include: 
(i) the interaction with electricity utilities and generators, \cite{Takc2021, Zhenhua2013}, 
(ii) more detailed power modeling of the DC facilities \cite{Radovanovic2022}, and 
(iii) the geographical location and load migration among a network of DCs \cite{Doyle2013}.

Within this large body of literature, the carbon-aware computing platform proposed in \cite{Radovanovic2023} stands out as it is 
currently implemented by Google to operate its inter-continental fleet of DCs. 
Google uses carbon intensity forecasts, and predicts future compute demand for their DCs. 
This information is used to generate so-called Virtual Capacity Curves (VCCs), which limit the hourly resource of their DCs in the fleet such that temporally flexible compute load is pushed to less carbon-intense hours.
Thus, by harnessing the compute load predictions, they can minimize carbon emissions.

In this paper, we design a novel day-ahead scheduling mechanism to allocate batch compute jobs over an interconnected fleet of DCs, inspired by Google's carbon-intelligent platform in \cite{Radovanovic2023}. 
The approach in \cite{Radovanovic2023} is scheduler-agnostic, meaning VCCs are computed independently of the actual compute job schedule. 
In contrast, we co-design the VCCs and compute job allocation using a hierarchical game-theoretic approach that distinguishes between separate individual compute jobs.
Additionally, our mechanism considers not only temporal shifting but also spatial migration.

The contributions of this paper are threefold. Firstly, we formalize the optimal allocation problem of compute jobs with spatial and temporal flexibility, over a network of DCs as a single-leader multiple-follower Stackelberg game. 
At the lower level, the owners of compute jobs compete for the computational resources in the DC network to process their jobs as soon as possible. 
At the upper level, the DC operator generates the virtual capacities to induce a competitive allocation that reduces carbon emissions and induces peak shaving.
Thirdly, we derive an efficient ad-hoc algorithm to solve the resulting large-scale Stackelberg game by adapting the method proposed in \cite{Grontas2023}, which begs similarities to the approach of solving a Stackelberg game in \cite{Maljkovic2023}. 
Finally, we numerically validate the proposed allocation mechanism using real carbon intensity data for a selection of Google's DC locations. 

Our numerical findings show that temporally shifting and spatially migrating flexible compute jobs significantly reduce carbon emissions. 
In contrast to scheduler-agnostic approaches, co-designing the VCCs can improve the homogeneity of waiting times in the job schedule.

\emph{Notation:}
$\mathbb{R}$, $\mathbb{R}_{\geq 0}$, $\mathbb{R}_{>0}$ denote the set of real, nonnegative real, and positive real numbers, respectively. 
Given $N$ scalars $a_1, \dots , a_N$, $\text{diag}(a_1,\dots , a_N )$ denotes the diagonal matrix with $a_1, \dots , a_N$ on the main diagonal.
Given $N$ column vectors $x_1,\dots, x_N  \in \mathbb{R}^n$,  $ x = \text{col}(x_1,\dots,x_N) = {[x^\top_1 ,\dots,x^\top_N]}^\top$ denotes their vertical concatenation. 
The Euclidean projection onto the set $\mc{X}$ is denoted as $\mathbb{P}_{\mc{X}} [\cdot]$. 
The partial derivative of a function $f$ with respect to its $i$-th argument is denoted as: $\nabla_i f(x_1, \dots, x_n ) = \dfrac{\partial f (x_1, \dots, x_n )}{\partial x_i}$. 

\section{PROBLEM STATEMENT}

We consider a set $\mc{I} := \{1,\dots,I\}$ of batch compute jobs to allocate across a fleet of DCs $\mc{D} := \{1, \dots, D \}$, over a planning horizon $\mc{T} := \{1, \dots, T \}$. 
Each batch job $i \in \mc{I}$ is uploaded by a customer, or team, at an initial location $d^i \in \mc D$, at which the data required for the computation is assumed to be physically stored. 
Each batch job is characterized by a predicted compute volume $v^i$ and a priority parameter $\tau^i$, describing its time urgency.
The DCs are physically interconnected via a fiber network that allows compute jobs to be transferred across the fleet. 
This network is modeled via a weighted graph $\mc{G}:= \mc{G}(\mc{D}, \mc{E})$, where the vertices are the DCs, and the edges are the fiber connections $e_{k,\ell} \in \mc{E}$ directly connecting neighboring DCs $k, \ell \in \mc{D}$. 
The weights of the edges encode the price of migrating the data, established by the internet service providers for using their routers. 

\subsection{Team's allocation game}
Each team shall choose the allocation of their batch job $i \in \mc{I}$, represented by the vector $y^i = \textrm{col}(y^i_1,\ldots,y^i_D)$, whose entries  $y^i_{d,t} \in \mathbb{R}_{\geq 0}$ describe the share of the compute volume $v^i$ allocated to DC $d \in \mc D$ at time $t\in \mc T$. 
In other words, $y^i$ determines where and when to reserve DC capacities to execute compute job $i$.
The allocation chosen by each team must satisfy various operational constraints. Intuitively, it must cover the total compute volume $v^i$, yielding
\begin{align}
    \sum_{t \in \mc{T}} \sum_{d \in \mc{D}} y^i_{d,t} = v^i, \quad \forall i \in \mc I. \label{eq:total_vi}
\end{align}

\begin{figure}[t]
	\centering
	\includegraphics[width=\linewidth]{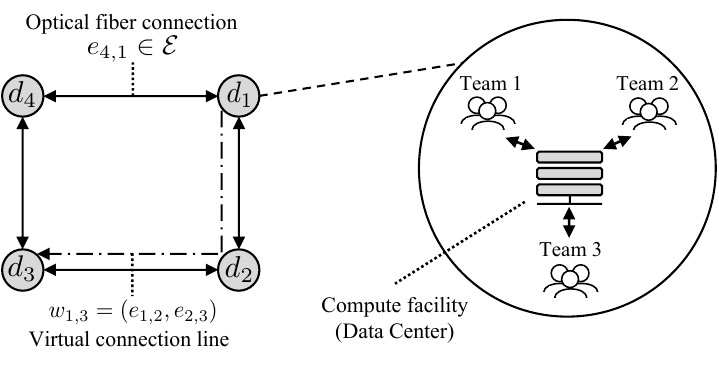}
	\caption{DC network featuring 4 DC locations interconnected by physical connection lines (solid line). A job migration between DC 1 and DC 3 is realized via a path through the fiber network. Right: Teams are associated with a DC location, where they initially submit their compute jobs.}
	\label{fig:dc_net_schematic}
\end{figure}

A job $i \in \mc I$ may be transferred from its initial DC $d^i$, to any alternative DC $j \in \mc{D}$ via a predetermined shortest path \( \omega_{i,j} \) over the fiber-optic network \( \mathcal{G} \). This path is uniquely identified and computed in advance, to ensure minimal transit time and resource usage, and consists of a sequence of fiber connections $e \in \mc{E}$. 
In Fig.~\ref{fig:dc_net_schematic}, we illustrate an example of the considered setup, where the path $\omega_{1,3} = (e_{1,2},e_{2,3})$ uses the fiber connections $e_{1,2}$ and $e_{2,3}$. 

Any migrated share of $v^i$ appears in a migration variable $z^i = \textrm{col}(z^i_1,\ldots,z^i_D)$, whose entries $z^i_{j,t} \in \mathbb{R}_{\geq 0}$ describe how much predicted load is migrated from $d^i$ to DC $j \in \mc{D}$ at time step $t \in \mc{T}$, over the path $\omega_{i,j}$. 
The entry $z^i_{i,t}$ describes the queue of unprocessed compute volume at this DC. 
Since there is no migration before the first time step, there are $T\!-\!1$ migration steps.
The share of a predicted compute volume $v^i$ executed at a DC other than the initial $d^i$, i.e., $d \in \mc{D}\setminus \{d^i\} $, must be migrated in the time step before
\begin{equation}
\label{eq:prior_migration}
\begin{gathered}
    y^i_{d, t+1} = z^i_{d,t}, \\ 
    \quad \forall i\in \mc{I}, \; \forall d \in \mc{D}\setminus \{d^i\}, \; \forall t \in \mc{T} \setminus \{T\}.
\end{gathered}
\end{equation}

Any share of a compute job $v^i$, that has not yet been computed at the initial DC $d^i$ or migrated to another DC $z^i_{j \neq i, \ell}$ is stored in the queue for the next time step $z^i_{i,t}$
\begin{equation}
\label{eq:queue}
\begin{gathered} 
    v^i - \sum_{\ell=1}^{t} \left( y^i_{d^i,\ell} + \sum_{j = 1, j \neq i}^D z^i_{j,\ell}  \right) =  z^i_{i,t},\\
    \forall i \in \mc{I}, \; \forall t \in \mc{T}\setminus \{T\}.
\end{gathered}
\end{equation}

Finally, the total amount of compute job volume allocable at each DC $d \in \mc D$ is limited by  
\begin{gather}
    \sum_{i \in \mc{I}} y^i_{d,t} \leq x_{d,t},  \quad \forall d \in \mc{D}, \; \forall t \in \mc{T} \label{eq:leq_VCC}, 
\end{gather}
where $x_{d,t}$ is a virtual capacity, set by the DC operator. 
These virtual capacities, considered over the complete planning horizon and all DCs in the network, constitute the VCCs.

Let $\bm{y}^i=\text{col}(y^i, z^i)$ denote the stacked vector of the local decision variables (namely, allocation and migration) of team~$i$. All local operational constraints \eqref{eq:total_vi}--\eqref{eq:leq_VCC} can be compactly represented via the set-valued mapping
\begin{align}
\label{eq:opconstr}
\mc Y^i(x,\bm{y}^{-i}) := \big\{ (y^i,z^i) ~|~ \eqref{eq:total_vi}-\eqref{eq:leq_VCC} \text{ hold} \big\},
\end{align} 
which depends on the VCCs $x=\textrm{col}(x_1,\ldots x_D)$ viewed as parameters, as well as the allocations of the other compute jobs $\bm{y}^{-i} = \textrm{col}(\bm{y}^{1},\ldots, \bm{y}^{i-1}, \bm{y}^{i+1}, \ldots, \bm{y}^I)$.


The objective of each team is to choose an allocation that minimizes execution time, migration cost, and deviation from a pre-determined allocation profile $\hat{y}^i$, formulated as
\begin{align}
\label{eq:CFi}
\begin{split}
    J^i(\bm{y}^i) &= \sum_{t \in \mc{T}}  \sum_{d \in \mc{D}} \left( \tau^i_t y^i_{d,t} 
+  z^i_{d,t} \left( \sum_{e \in \omega_{i,d}} \sigma^i_e \right)\right) \\ 
& \textstyle \quad + \frac{1}{2} \epsilon \|\bm{y}^i-\hat{\bm{y}}^i \|^2.
\end{split}
\end{align}

The first term in \eqref{eq:CFi} penalizes allocating compute jobs incrementally with the time steps and depends on the priority parameter $\tau^i_t = {\tau}^i l_t$, where $l_t = t/T$ is a time-dependent weight that penalizes delayed allocations.
The second term in \eqref{eq:CFi} penalizes the migration of compute jobs, as transferring data over the network comes with a time penalty for each team.
Therein, the total price of migration is the sum of all the prices at each fiber connection $\sigma^i_e = \tau^i \sigma_e$, multiplied by the priority parameter to further penalize migrations of urgent compute jobs.
The quadratic terms penalize the deviation from a predefined allocation and migration $\bm{\hat{y}}^i = \text{col}(\hat{y}^i, \hat{z}^i)$. 

Overall, each team solves the following optimization problem to find their optimal allocation
    \begin{align}
    \label{eq:cond_game}
  (\forall i \in \mc I): \quad      \min_{\bm{y}^i} \; J^i(\bm{y}^i) \quad  \text{s.t.} \quad \bm{y}^i \in \mc Y^i(x,\bm{y}^{-i}).
    \end{align}
The collection of these inter-dependent optimization problems constitutes a generalized game \cite{Belgioioso2022}, parametric in $x$, the VCCs.
Note that the other teams' allocations $\bm{y}^{-i}$ enter the local constraints in \eqref{eq:cond_game} due to the resource constraint \eqref{eq:leq_VCC}, rendering the game a generalized game. 
Furthermore, with the teams' objectives in \eqref{eq:CFi} being decoupled, this game structure corresponds to an exact potential game \cite{Facchinei2011}.

A meaningful solution concept for \eqref{eq:cond_game} is the Generalized Nash Equilibrium (GNE), which is a set of allocations $\bar{\bm y} = (\bar{\bm y}^1, \ldots, \bar{\bm y}^I )$ that simultaneously solve the optimization problems in \eqref{eq:cond_game}.
Here, we focus on the special subclass of variational GNEs (v-GNEs) due to their computational tractability and economic fairness \cite{Belgioioso2022}. This subclass of equilibria corresponds to the solution set of the parameterised variational inequality $\text{VI}(F,\mc{Y}(x))$, namely, the problem of finding a vector $\bar{\bm{y}}$ such that
\begin{align}
\label{eq:VI}
\langle F(\bar{\bm{y}}), (\bm{y}-\bar{\bm{y}}) \rangle \geq 0, \quad \forall \bm{y} \in \mc{Y}(x),
\end{align}
where, $F(\bm{y}) := \text{col}(\{ \nabla  J_i(\bm{y}^i)\}_{i \in \mc{I}})$ is the \textit{pseudo-gradient} mapping of the game \eqref{eq:cond_game}, and $\mc Y(x)$ collects the operational constraints \eqref{eq:opconstr} and depends explicitly on the VCCs $x$. 
We denote by $\bm{y}_\star(\cdot)$ the parameter-to-solution mapping that, given the VCCs $x$, returns the set of solutions $\bm{y}_\star(x)$ to the VI in \eqref{eq:VI}, namely, a set of strategically-stable allocations.
It can be shown that $\bm{y}_\star(x)$ is single-valued for any $x$ that makes \eqref{eq:VI} feasible, meaning that the allocation game has a unique solution for given VCCs, as the pseudo-gradient mapping $F$ is strongly monotone.

\subsection{Supervisory Objectives}
\begin{figure}
	\centering
	\includegraphics[width=\linewidth]{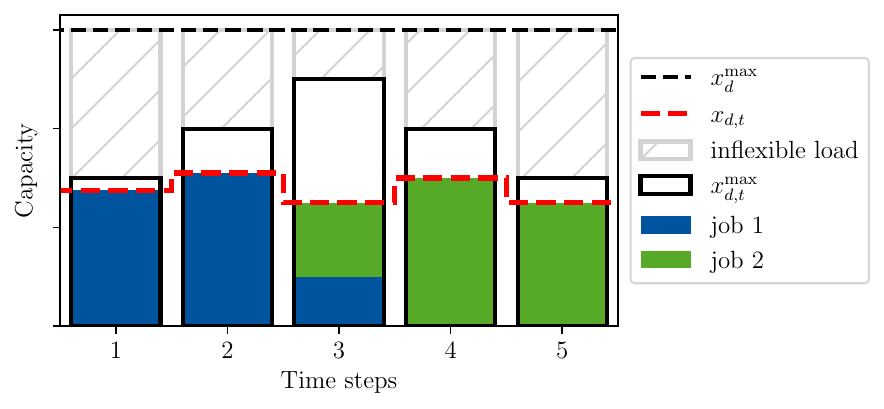}
	\caption{An example of sequential allocation of two compute jobs $y^1$ and $y^2$ on a data center $d \in \mc{D}$, where job 1 is of higher priority than job 2. The virtual capacity curve (red dashed line) limits the allocable load at each time slot. The maximum capacity of the DC $x^{\max}_{d,t}$ (solid black line) is obtained by subtracting the inflexible load from the maximum capacity of that DC $x^{\max}_d$ (dashed black line). }
	\label{fig:vcc_schematic}
\end{figure}

The DC operator can influence the outcome of the allocation game \eqref{eq:cond_game} by manipulating the virtual capacities $x_{d,t} \in \mathbb{R}_{\geq 0}$, limiting the total load at each DC and in each time step.
As for the allocations, also the VCCs must satisfy a series of operational constraints.
Firstly, the total available capacity defined by the VCCs must accommodate the total compute demand, yielding
\begin{align}
    \sum_{d \in \mc{D}} \sum_{t \in \mc{T}} x_{d,t} &\geq \sum_{i \in \mc{I}} v^i. \label{eq:leader_total} 
\end{align}
Additionally, the VCCs cannot exceed the physical computational capacity of the DCs, yielding
\begin{align}
	0 \leq x_{d,t} &\leq x^{\max}_{d,t},   \quad \forall d \in \mc{D}, \; t \in \mc{T}, \label{eq:X_1}
\end{align}
where the maximum capacity is denoted by $x_{d,t}^{\max}$. 
This value is obtained by subtracting the inflexible demand, which we cannot shift or migrate but assume to be forecast perfectly, from the physical capacity $x_d^{\max}$.
Fig. \ref{fig:vcc_schematic} displays an exemplary job schedule for two jobs $y^1$ and $y^2$, with the VCC $x_{d,t}$ as their upper bound and physical constraint $x^{\max}_{d}$.
The operational constraints on the DC operator's decision are compactly represented via
\begin{align}
\label{eq:DC_opconstr}
\mathcal{X} := \big\{ x  ~|~ \eqref{eq:leader_total}, \eqref{eq:X_1} \text{ hold} \big\}.
\end{align} 

The DC operator utilizes its influence on the teams' allocations to reduce the carbon impact and peak usage of the DCs. 
This goal is modeled by the multi-objective function
\begin{align}
	\label{eq:ul_cost_general}
	\phi (x, \bm{y}) &=  \phi^{\text{carb}}( \bm{y}) + \phi^{\text{peak}}( \bm{y}) + \xi \phi^{\text{migr}}( \bm{y}).
\end{align}
The term $\phi^\text{carb}$ penalizes carbon emissions and is defined as
\begin{align}
	\phi^{\text{carb}}(\bm{y}) &=  \sum_{d \in \mc{D}} \; \sum_{t \in \mc{T}} \; \rho^\text{carb}_{d,t}\left( \sum_{i \in \mc{I}} \;  y^i_{d,t}\right),
\end{align}
where $\rho^{\text{carb}}_{d,t}$ is the carbon intensity of the supplied power at DC $d$ and time step $t$, modeling the carbon impact of utilizing computing power in a linear relationship, as it is done in \cite{Minxian2020, Tran2016}. 
The second term $\phi^\text{peak}$ penalizes peak usage of DCs computational resources per time step and is defined as
\begin{align}
    \label{eq:peak_shaving}
	\phi^{\text{peak}}(\bm{y}) &= \sum_{d \in \mc{D}} \left( \sum_{t \in \mc{T}} \left( \sum_{i \in \mc{I}} y^i_{d,t} \right)^p \right)^{1/p},
\end{align}
for some $p \in \mathbb{N}$ large enough to approximate the infinity norm. The third term penalizes the aggregate migration of compute jobs and is defined as
\begin{align}
	\phi^{\text{migr}}(\bm{y}) & = \sum_{i \in \mc{I}} \sum_{t \in \mc{T}} \sum_{d \in \mc{D}} z^i_{d,t} \left( \sum_{e \in \omega_{i,d}} \sigma^i_e \right) .
\end{align}
This last term is necessary to reduce network traffic from the DC operator's perspective, as the DC operator is responsible for paying network routing fees to internet service providers. 
Finally, the parameter $\xi \geq 0 $ in \eqref{eq:ul_cost_general} regulates the impact of migration on the final allocation.
In contrast to an exclusively monetary objective, the multi-objective in \eqref{eq:ul_cost_general} addresses sustainability aspects directly.  

Overall, the DC operator aims to solve the following single-leader multiple-follower Stackelberg game
\begin{subequations}
\label{eq:bilevel_game}
\begin{align}
		\min_{x, \bm{y}} \quad &\phi(x, \bm{y}) \label{eq:cF} \\
		\text{s.t.   } \quad &x \in \mc{X} \\
		\quad &
\langle F({\bm{y}}), (\bm{y}'-{\bm{y}}) \rangle \geq 0, \quad \forall \bm{y}' \in \mc{Y}(x).
		\label{eq:EQC}
\end{align}
\end{subequations}

\section{BILEVEL GAME SOLUTION APPROACH}
We solve \eqref{eq:bilevel_game} using a customized version of a first-order algorithm named BIG Hype, recently proposed in \cite{Grontas2023}. 

\subsection{A Hyper-gradient based algorithm}
The first-order algorithm in \cite{Grontas2023} is based on the idea of substituting the equilibrium constraints \eqref{eq:EQC} into the objective function \eqref{eq:cF}, by exploiting the solution mapping $\bm{y}_\star(\cdot)$ and the fact that solutions to \eqref{eq:VI} are unique.
The resulting non-convex non-smooth optimization problem reads as
\begin{align}
	\begin{split}
		\min_x \quad & \phi (x, \boldsymbol{y}_\star(x)) =: \phi_e(x) \label{eq:bilevel_single_level}\\
		\text{s.t.} \quad & x \in \mc{X}.
	\end{split}
\end{align}
Then, a local solution to \eqref{eq:bilevel_single_level} is obtained by relying on projected ``gradient" descent.
Whenever $\bm{y}_\star(x)$ is differentiable at $x$, one can obtain a gradient by applying the chain rule\footnote{If not differentiable, the standard Jacobians are replaced by elements of the conservative Jacobians. For detailed explanations and proofs, see \cite{Grontas2023}.}
\begin{gather}
    \nabla \phi_e(x) = \nabla_1 \phi( x, \boldsymbol{y}_\star(x)) + \boldsymbol{J y}_\star(x)^\top \nabla_2 \phi(x,\boldsymbol{y}_\star(x)), \label{eq:hypergrad}
\end{gather}
where $\boldsymbol{J}\boldsymbol{y}_\star(x)$ is the Jacobian of the solution mapping $\boldsymbol{y}_\star(\cdot)$ at $x$, commonly known as the sensitivity.

The proposed algorithm is summarized in Algorithm~\ref{alg:hypergradient_general} and consists of three main steps. 
In step \textbf{1}, the DC operator uses the current estimates of the equilibrium and sensitivity to compute a hypergradient \eqref{eq:hypergrad}, and runs a projected hypergradient descent step.
Based on the new VCCs $x^{k+1}$, in step \textbf{2}, the teams compute the resulting optimal allocation $\boldsymbol{y}_\star(x^{k+1})$ and subsequently, in step \textbf{3}, its sensitivity $\boldsymbol{J}\boldsymbol{y}_\star(x)$.

\SetKwProg{myproc}{repeat}{}{}
\begin{algorithm}
	Initialize $x^0 \in \mc{X}$, $\bm{y}^0 \in \mathbb{R}^{n_y}_{\geq0}$, $s^0 = \bm{0} \in \mathbb{R}^{n_y\times n_x}$, $k = 1$, and  $\{\alpha^k\}_{k \in \mathbb{N}}$ \\
	\myproc{\emph{until convergence}}{ 
	\BlankLine
	\textbf{1}. DC operator's projected hypergradient step: \\
	$\quad  \nabla \phi_e^k \leftarrow \nabla_1 \phi( x^k, \bm{y}^k) + (s^k)^\top \nabla_2 \phi(x^k,\bm{y^k}) $\\
	$\quad  x^{k+1}\leftarrow \mathbb{P}_\mc{X} [x^k - \alpha^k \nabla \phi_e^k] $\\
	\BlankLine
	\textbf{2}. Equilibrium seeking step: \\
	$\quad \bm{y}^{k+1}\leftarrow \bm{y}_\star (x^{k+1})$ \\
	\BlankLine
	\textbf{3}. Sensitivity computation step: \\
	$\quad  s^{k+1} \leftarrow \bm{J}\bm{y}_\star(x^{k+1})$ \\
}
\caption{Customized BIG Hype} \label{alg:hypergradient_general}
\end{algorithm}

Under the considered problem setup, the convergence of Algorithm~\ref{alg:hypergradient_general} to critical points of \eqref{eq:bilevel_single_level} follows by Theorem~2 in \cite{Grontas2023}. We omit the details here for the sake of brevity.
In the following subsections, we describe how to perform steps \textbf{2} and \textbf{3} of Algorithm~\ref{alg:hypergradient_general} for the specific game setup in \eqref{eq:cond_game} in a computationally efficient manner.

\subsection{Equilibrium Seeking Step}
\label{subsec:eq_comp}
To efficiently find a solution of the allocation game \eqref{eq:cond_game}, we consider the surrogate optimization problem
\begin{subequations}%
\label{eq:miao}
\begin{align}
\min_{\bm{y}} &\quad \sum_{i \in \mc I} J^i(\bm{y}^i) \label{eq:miao_obj}\\
& \quad  \text{s.t.} \quad \bm{y} \in \mc Y(x), \label{eq:miao_constr}
\end{align}
\end{subequations}
whose minimizer corresponds to the v-GNE of \eqref{eq:cond_game}.
The individual objective functions \eqref{eq:CFi} are decoupled, such that the sum of all teams' objectives constitutes the potential function \cite[p. 243]{Facchinei2011}, allowing us to rewrite \eqref{eq:cond_game} as \eqref{eq:miao}.
A formal proof can also be obtained by comparing the Karush-Khun-Tucker (KKT) conditions \cite[Th. 4.8]{Facchinei2010}, but is omitted here due to space limitations. 
The sum in \eqref{eq:miao_obj} can be written as
\begin{align}
    J(\bm{y}) = \sum_{i \in \mc{I}} J^i(\bm{y}^i) = q^\top \bm{y} + \frac{1}{2}\epsilon (\bm{y}-\bm{\hat{y}})^\top (\bm{y}-\bm{\hat{y}}), 
\end{align}
while the collection of constraints \eqref{eq:miao_constr} can be expressed as
\begin{align}
    \mc{Y}(x) = \left\{ \bm{y} \in \mathbb{R}^{n_y} | A \bm{y} = b, G \bm{y} \leq h + Hx \right\}. \label{eq:operational_constr_total}
\end{align}
To improve the computational efficiency, we reduce the problem size by eliminating variables, as outlined in \cite[Sec. 10.1.2]{Boyd2004}. 
We substitute $\bm{y} = F_T \bm{\Tilde{y}} + \bm{y}^\dagger$ into $J(\bm{y})$, where the transformation matrix $F_T$ consists of basis vectors of the nullspace of $A$ and $\bm{y}^\dagger$ is any solution to $A\bm{y}^\dagger = b$. 
By choosing $\bm{\hat{y}} = \bm{y}^\dagger$, the objective reduces to $J(\bm{\Tilde{y}}) = (F_T q)^\top \bm{\Tilde{y}} + \frac{1}{2} \epsilon \bm{\Tilde{y}}^\top F_T^\top F_T\bm{\Tilde{y}}$. 
The new optimization variable $\bm{\Tilde{y}}$ is reduced in dimension by the number of rows in $A$, i.e., the number of equality constraints.
Finally, the optimization problem reads as follows
\begin{align}
	\begin{split}
		\label{eq:final_qp}
		\min_{\tilde{\bm{y}}} & \quad \|  F_T \bm{\tilde{y}} +{q}/ \epsilon \|_2^2 \\
		\text{s.t. } & \quad \tilde{G} \bm{\tilde{y}} \leq \tilde{h} + H x,
	\end{split}
\end{align}
where $q$,  $\Tilde{G} = G F_T$, $\Tilde{h} = h - G \bm{y}^\dagger$, ${H}$ depend on the problem data. 
Problem \eqref{eq:final_qp} is an inequality-constrained quadratic program (QP) with a convex objective, that can be efficiently solved using off-the-shelf solvers. 

\subsection{Sensitivity Computation}
The equilibrium sensitivity $\bm{J y_\star}(x)$ corresponds to the sensitivity of the solutions to the QP \eqref{eq:final_qp} with respect to changes in $x$, which we compute using the approach in \cite{AmosYear}. 
First, we find the total differentials of the KKT conditions of \eqref{eq:final_qp}, more specifically of the stationarity condition and complementarity slackness at the optimal point
\begin{align}
    0 &= \Tilde{G}^\top d \bm{\lambda} + d \bm{\tilde{y}},\\
    0 &= \text{diag}(d\bm{\lambda})
 ( \Tilde{G} \bm{\tilde{y}_\star} - \tilde{h} - Hx) + \text{diag}(\bm{\lambda_\star})(\Tilde{G} d \bm{\Tilde{y}} - H dx), 
\end{align}
with the Lagrange multipliers $\bm{\lambda}$ for the inequality constraints and the differentials $d\boldsymbol{\lambda}$, $d\bm{\Tilde{y}}$ and $dx$. 
The columns of the sensitivity matrix $\bm{J y_\star}(x)$ are the solutions of $d\bm{\Tilde{y}}$ when replacing $dx$ by the columns of the identity matrix.
Mathematically, for each single constraint, the optimal solution $\bm{\Tilde{y}_\star}$ and the differential $d\bm{\lambda}$ do not influence $d\bm{\Tilde{y}}$ and $dx$, because either $\Tilde{G} \bm{\tilde{y}_\star} - \tilde{h} - Hx = 0 $ (active constraint), or $\bm{\lambda}_\star~=~0$ (inactive constraint), or both are zero.
Intuitively, since the DC operator's decisions only influence the lower level through the constraints, the sensitivity does not depend on the teams' objectives. 
Therefore, we only consider the rows of $\Tilde{G}$ and $H$ which correspond to the active inequality constraints and solve the following system of linear equations
\begin{align}
	\tilde{G}_{k,\cdot} \; d\bm{\tilde{y}} - H_{k,\cdot} \; dx = 0, \quad \forall k \text{ with } \lambda_{\star,k} > 0.\label{eq:underdet_lse}
\end{align}

\section{SIMULATION RESULTS}

\subsection{Simulation Setup}

We consider a network of 12 DCs distributed over 4 continents, whose locations we source from Google's fleet of DCs \cite{DCLocations2023}. 
We use real carbon intensity data for those locations, obtained from Electricity Maps \cite{Emaps2023}, which covers 24 hours with a sampling rate of 5 hours, starting on February 22\textsuperscript{nd}, 2023, at 10 o'clock.
The maximum computational capacities $x^{\max}_{d,t}$ of the DCs vary due to inflexible loads that fluctuate in a sinusoidal shape, similar to \cite{Hogade2022}. 
We set the parameter for the quadratic penalty in the team's objective to a small value $\epsilon = 2\times10^{-8} $, which means that teams virtually have no preferred allocation, but only care about the processing time and migration cost.
The $p$-norm parameter for the peak shaving cost in \eqref{eq:peak_shaving} is set to $p=6$, a value approximating the infinity norm while retaining numerical stability.
To improve the conditioning of the DC operator's cost function \eqref{eq:ul_cost_general}, we include the uniform term $\frac{1}{2} \mbf{1}_{n_x}^\top x $, that ensures that $\nabla_1 \phi(x, \bm{y_\star}(x))$ is nonzero. 
Moreover, to ensure the feasibility of the parametrized allocation game \eqref{eq:cond_game}, at each step of Algorithm \ref{alg:hypergradient_general}, we impose the following: In the first time step, the VCCs at each DC shall not exceed the cumulative predicted load of all teams, that have their data uploaded at this DC, i.e.,  $x_{d,1} \leq \sum_{ \{i \in \mc{I} \; | \; d = d^i\}} v^i$, for all $  d \in \mc{D}$.

\subsection{The Impact of Temporal Shifting \& Spatial Migration}

\begin{figure}[t]
	\begin{minipage}[t]{0.48\linewidth}
    \includegraphics[width=\linewidth]{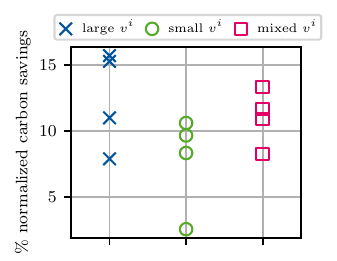}
    \caption[Carbon emission savings due to time shifting.]{Carbon emission savings normalized by compute volume due to time-shifting. Bilevel game vs, na\"ive approach (full capacity utilization) in three scenarios.} 
	\label{fig:timeshift_savings}
    \end{minipage}
 \hspace*{\fill}
	\begin{minipage}[t]{0.48\linewidth}
    \includegraphics[width=\linewidth]{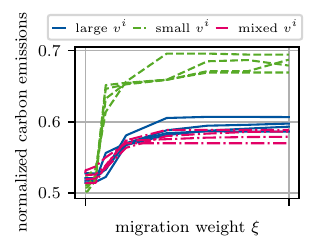}
	\caption[Investigation of migration weight.]{Carbon emissions, normalized by compute volume, vs migration price $\xi$, for three scenarios. The growth demonstrates a positive impact of spatial migration.}
	\label{fig:migration_large}
	\end{minipage}
\end{figure}

To analyze the impact of temporal shifting, we compare our allocation mechanism with a na\"{i}ve algorithm that allocates the compute jobs according to their priority, always using the full available capacity $x_{d,t}^\text{max}$. 
We consider three load scenarios, each characterized by different compute job volumes: 
(a) large compute jobs that typically require more than one time step at a DC location; (b) multiple small compute jobs that can be processed during a single time step; and (c) a mixture of large and small compute jobs.
As shown in Fig. \ref{fig:timeshift_savings}, temporally shifting compute loads significantly reduces carbon emissions, depending on the load scenario. 

To investigate the impact of spatial migration on carbon emissions, we modify the migration price $\xi$ in the DC operator's objective function.
A very large $\xi$ corresponds to disabling the migration of compute jobs. 
As shown in Fig. \ref{fig:migration_large}, increasing $\xi$ increases carbon emissions. 
This demonstrates the potential to save carbon emissions that comes with enabling spatial migration.

\subsection{Co-design vs. Sequential Optimization} In this case study, we compare our bilevel approach for co-designing VCCs and allocations with a sequential optimization approach, similar to the scheduler-agnostic scheme used by Google \cite{Radovanovic2023}. The latter consists of two sequential steps:
In the first step, optimal VCCs are computed by solely using a forecast of the compute load and carbon intensity. In the second step, the teams find the optimal allocation given the previously computed VCCs.
In this sequential approach, the DC operator must optimize over the worst-case allocation scenarios, namely, the VCCs, rather than the actual allocations. The sequential approach does not take the migration of compute loads into account. Thus, the resulting cost function of the DC operator reads as
\begin{align}
\hat \phi(x) =	 \sum_{d \in \mc{D}}  \sum_{t \in \mc{T}} \rho^\text{carb}_{d,t} x_{d,t} + \sum_{d \in \mc{D}} \left( \sum_{t \in \mc{T}} ( x_{d,t} )^p \right)^{1/p},
\end{align}
where the actual $\sum_{i \in \mc{I}} y^i_{d,t}$ have been replaced by the VCCs $x_{d,t}$. Overall, this sequential optimization scheme reads as
\begin{align}
	\text{Step 1: }\quad& \bar{x} = \arg \min_x \; \hat \phi(x) ,\\
	 \text{Step 2: }\quad & \bm{y} = \bm{y}_\star(\bar x).
	\label{eq:SQ-OP}
\end{align}

Further, we introduce the following two metrics to evaluate the performance of the allocations:
\begin{enumerate}
	\item allocation fairness: $\psi \left(\left\{\tau^{\text{time},i} (y^i) / v^i \right\}_{i \in \mc{I}}\right)$,
	\item total waiting time: $\sum_{i \in \mc{I}} \tau^{\text{time},i} (y^i)$, 
\end{enumerate}
where $\psi(\cdot) $ is the empirical standard deviation and $\tau^{\text{time},i}(y^i)$ denotes the time cost, i.e., the first linear term in \eqref{eq:CFi}. The fairness criterion represents the heterogeneity of the waiting times among the teams. 
In contrast, the second criterion is the total wait time for all jobs, weighted by their time priority parameter $\tau^i$.
\begin{figure}[t]
    \centering
    \includegraphics[width=\linewidth]{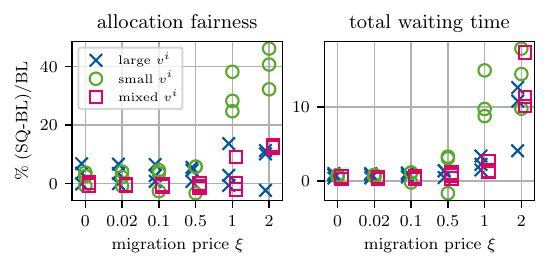}  
    \caption{Fairness and total waiting time in a direct comparison between our approach (BL) and sequential optimization (SQ). Sequential optimization scores higher (worse) values in both, and the difference grows with increasing migration price.} 
    \label{fig:fairness_welfare}
\end{figure}

\begin{figure}[t]
    \centering
    \includegraphics[width=\linewidth]{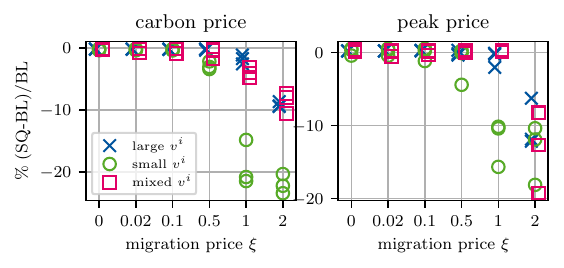}  
    \caption{Carbon emissions and peak price in a direct comparison between our approach (BL) and sequential optimization (SQ). SQ scores lower (better) values in both for high migration prices in all three scenarios.} 
    \label{fig:carbon_peak}
\end{figure}

In Fig. \ref{fig:fairness_welfare}, we show the outcome of simulations with varying migration prices to showcase the differences between the bilevel game \eqref{eq:bilevel_game} and the sequential approach \eqref{eq:SQ-OP}.
Generally, fairness and waiting time are improved when co-designing VCCs and the compute job schedule. The difference becomes even more apparent with higher values of $\xi$, as this directly addresses the shortcoming of not featuring migration in the sequential approach. 
However, as demonstrated by the results in Fig.~\ref{fig:carbon_peak}, the sequential optimization approach scores lower carbon emissions and better peak-shaving performance for large migration prices.
In summary, if we enable spatial shifting, the co-design of VCCs and allocations allows us to perform similarly to the sequential approach regarding decarbonization and peak-shaving. Yet, co-design can lead to reduced and fairer waiting times for the teams. This may incentivize users to participate in the proposed coordination mechanism.

\section{Conclusions}

We modeled the problem of co-designing the allocation of flexible batch compute jobs and virtual capacity curves of DCs as a bilevel game. This formulation promotes the supervisory control objectives of the DC operator and models each team as a player competing for the computational resources in the DCs.
A local solution of the resulting game is found by deploying the recently developed algorithm \cite{Grontas2023}, with some ad-hoc modifications that improve efficiency.
Simulation results show that allowing for spatial migration and temporal shifting of compute jobs has a high potential to reduce carbon emissions in the operation of DCs. 
Moreover, compared with standard sequential optimization, our hierarchical approach can reduce the total waiting time of compute jobs and improve the fairness of their allocation.

\balance

\addtolength{\textheight}{0cm}   




%


\bibliographystyle{IEEEtran}
\bibliography{IEEEabrv,mybibfile}

\end{document}